\documentclass[aps,prb,showpacs,twocolumn]{revtex4}
\usepackage{graphicx}
\begin{document}
\title{Current-induced magnetoresistance oscillations
in two-dimensional electron systems}
\author{X. L. Lei}
\affiliation{Department of Physics, Shanghai Jiaotong University,
1954 Huashan Road, Shanghai 200030, China}

\begin{abstract}
Electric current-induced magnetoresistance oscillations recently discovered
in two-dimensional electron systems are analyzed using a microscopic 
scheme for nonlinear magnetotransport 
direct controlled by the current. The magnetoresistance oscillations are shown 
to result from drift-motion assisted electron scatterings between Landau levels. 
The theoretical predictions not only reproduce all the main features   
observed in the experiments but also disclose other details 
of the phenomenon. 

\end{abstract}


\maketitle

The effect of a strong dc on magnetoresistance has long been 
an outstanding problem in transport in two-dimensional (2D) electron systems (ESs).
In the case of microwave-induced magnetoresistance 
oscillations,\cite{Ryz,Zud01,Ye,Mani,Zud03,Dor03,Durst} 
a finite dc has been shown to suppress the oscillation and eliminate 
the negative resistance existing in the weak current limit.\cite{Lei03,Ng}
Quite surprisingly, in the case without microwave, a relatively weak dc can 
induce substantial magnetoresistance oscillations in 2DESs.   

The current-induced magnetoresistance oscillations (CIMOs) 
were observed in differential magnetoresistance 
in high-mobility Hall-bar specimens.\cite{Yang2002}
The oscillation is periodic in inverse magnetic field and its period  
is tunable by the current density.
This discovery was later confirmed in highly doped samples 
and the differential resistance oscillating 
with changing current density at fixed magnetic field 
was also detected.\cite{Bykov} Measurements 
were recently reported at higher temperatures\cite{JZ0607741} and  
careful studies in ultrahigh mobility samples were carried out.\cite{WZ0608727}

In this Letter we show that all these observed CIMOs can be well explained with the 
microscopic balance-equation scheme for hot-electron magnetotransport 
direct controlled by the current.\cite{Ting,Lei85,Lei851,Lei852}

We consider a 2D system consisting of $N_{\rm s}$ electrons in a unit area of 
the $x$-$y$ plane. These electrons, interacting with each other, are scattered by
random impurities and by phonons in the lattice.
There are a uniform magnetic field ${\bf B}=(0,0,B)$ along the $z$ direction
and a uniform electric field ${\bf E}$ in the $x$-$y$ plane.
The nonlinear steady state magnetotransport of this system can be described 
in terms of the center of mass and relative electron variables\cite{Ting,Lei85,Lei851}
by the following force and energy-balance equations:\cite{Lei852}
\begin{eqnarray}
N_{\rm s}e{\bf E}+N_{\rm s} e ({\bf v} \times {\bf B})+
{\bf f}({\bf v})&=&0,\label{eqforce}\\
{\bf v}\cdot {\bf f}({\bf v})+ w({\bf v})&=&0. \label{eqenergy}
\end{eqnarray}
Here ${\bf f}({\bf v})={\bf f}_{\rm i}({\bf v})+{\bf f}_{\rm p}({\bf v})$ 
is the frictional forces due 
to impurity and phonon scatterings, with the impurity part
\begin{equation}
{\bf f}_{\,\rm i}({\bf v})=\sum_{{\bf q}_\|} {\bf q}_\| \left| U({\bf q}_\|)\right| ^{2}
\Pi _{2}({\bf q}_\|,{\bf q}_\|\cdot {\bf v}), \label{forcei}\\
\end{equation}
\begin{equation}
w({\bf v})=\sum_{{\bf q}} \Omega_{\bf q} \left| M({\bf q})\right|^{2}\, \Lambda_{2}({\bf q},
\Omega _{{\bf q}}+{\bf q}_\|\cdot {\bf v}) 
\label{energy}
\end{equation}
is the electron energy-loss rate to the lattice.
In these expressions,
$U({\bf q}_\|)$ and $M({\bf q})$ stand for effective impurity and phonon
scattering potentials, 
$\Pi_2({\bf q}_\|,\Omega)$ is the imaginary part of the electron density-correlation function 
at electron temperature $T_{\rm e}$ in the magnetic field, and function
$
\Lambda_2({\bf q},\Omega)\equiv 2\,\Pi_2({\bf q}_\|,\Omega)
[n(\Omega_{\bf q}/T)-n(\Omega/T_{\rm e})]
$\, ($n(x)\equiv 1/({\rm e}^x-1)$), also relevant to phonon emission and absorption.
The effect of interparticle Coulomb interactions is included in the $\Pi_2$ function 
to the degree of electron level broadening and screening. 
The remaining $\Pi_2({\bf q}_{\|}, \Omega)$ function is that of a 2D electron gas  
in a magnetic field, which can be written in the Landau representation as:\cite{Ting}
\begin{eqnarray}
&&\hspace{-0.7cm}\Pi _2({\bf q}_{\|},\Omega ) =  \frac 1{2\pi
l_{\rm B}^2}\sum_{n,n'}C_{n,n'}(l_{\rm B}^2q_{\|}^2/2) 
\Pi _2(n,n',\Omega),
\label{pi_2q}\\
&&\hspace{-0.7cm}\Pi _2(n,n',\Omega)=-\frac2\pi \int d\varepsilon
\left [ f(\varepsilon )- f(\varepsilon +\Omega)\right ]\nonumber\\
&&\,\hspace{2cm}\times\,\,{\rm Im}G_n(\varepsilon +\Omega)\,{\rm Im}G_{n'}(\varepsilon ),
\label{pi_2ll}
\end{eqnarray}
where $l_{\rm B}=\sqrt{1/|eB|}$ is the magnetic length,
$
C_{n,n+l}(Y)\equiv n![(n+l)!]^{-1}Y^l{\rm e}^{-Y}[L_n^l(Y)]^2
$
with $L_n^l(Y)$ the associate Laguerre polynomial, $f(\varepsilon
)=\{\exp [(\varepsilon -\mu)/T_{\rm e}]+1\}^{-1}$ is the Fermi 
function at electron temperature $T_{\rm e}$, 
and ${\rm Im}G_n(\varepsilon )$ is the density of states (DOS) of the broadened LL $n$.

We model the DOS function 
with a Gaussian-type form for both overlapped and separated 
LLs ($\varepsilon_n=n\omega_c$ is the center energy of the $n$th LL and
$\omega_c=eB/m$ is the cyclotron frequency):\cite{Ando82}
\begin{equation}
{\rm Im}G_n(\varepsilon)=-(2\pi)^{\frac{1}{2}}\Gamma^{-1}
\exp[-2(\varepsilon-\varepsilon_n)^2/\Gamma^2]
\label{Gauss}
\end{equation}
with a magnetic field $B^{1/2}$ dependent half width
 $\Gamma=(8\alpha e\omega_c/\pi m \mu_0)^{1/2}$ expressed
in terms of $\mu_0$, the linear mobility at lattice temperature $T$ 
in the absence of the magnetic field, together with a broadening parameter $\alpha$ 
to take account of the difference between the transport scattering time 
and the broadening-related quantum lifetime.\cite{Mani,Durst}
We will also use a $B$-independent $\Gamma$ for comparison. 

Equations (\ref{eqforce}) and (\ref{eqenergy}) are quite general, 
applicable to current-control magnetotransport in any configuration. 
For an isotropic system where the frictional force is in the opposite direction of 
the drift velocity, we can write ${\bf f}({\bf v})=f(v){\bf v}/v$ and 
$w({\bf v})=w(v)$.  
In the Hall configuration with the velocity ${\bf v}$ in the $x$ direction
${\bf v}=(v,0,0)$ or the current densities $J_x=N_{\rm s}ev$ and $J_y=0$,
Eq.\,(\ref{eqforce}) yields the transverse and longitudinal resistivities 
\begin{eqnarray} 
&&R_{xy}= E_y/J_x=B/N_{\rm s}e,\label{eqrxy}\\  
&&R_{xx}= E_x/J_x=-f(v)/(N_{\rm s}^2e^2v),\label{eqrxx}
\end{eqnarray}
and the longitudinal differential resistivity
\begin{equation}
r_{xx}=-({\partial f(v)}/{\partial v})/(N_{\rm s}^2e^2). \label{eqdr}
\end{equation}
Equation (\ref{eqrxy}) confirms the relation 
$E_y=BJ_x/N_{\rm s} e$ in the Hall configuration of nonlinear magnetotransport 
even with an intense dc flowing in the $x$ direction. 

Expressions (\ref{eqrxx}) and (\ref{eqdr}) are particularly convenient  
to deal with current-induced phenomena. 
Apparently, the velocity $v$, or the current $J_x$, can affect $R_{xx}$ and $r_{xx}$ 
through the ${\bf q}_\|\cdot {\bf v}$ factor in the 
$\Pi_2({\bf q}_\|,{\bf q}_\|\cdot {\bf v})$ function.
Eqs.\,(\ref{pi_2q}) and (\ref{pi_2ll}) indicate 
that in the case of low electron temperature ($T_{\rm e}\ll \epsilon_{\rm F}$, 
the Fermi level) and many Landau-level occupation,
$\Pi_2({\bf q}_\|, \Omega)$ is essentially a periodical function, i.e.
$\Pi_2({\bf q}_\|, \Omega+\omega_c)=\Pi_2({\bf q}_\|, \Omega)$.
Therefore, the impurity-induced resistivity $R_{xx}$ would exhibit periodical 
oscillations
when changing drift velocity $v$ or changing magnetic field $1/B$.
We introduce a frequency-dimension quantity $\omega_j\equiv 2k_{\rm F}v$  to 
trace the change of the drift velocity $v$ or the current $J_x=N_{\rm s} e v$,
and use the dimensionless ratio 
\begin{equation}
\frac{\omega_j}{\omega_c}=\frac{2mk_{\rm F}v}{eB}
=\sqrt{\frac{8\pi}{N_{\rm s}}}\frac{m}{e^2}\frac{J_x}{B}
\end{equation}
as the control parameter to demonstrate this oscillation, which  
exhibits an approximate periodicity $\Delta(\omega_j/\omega_c)\sim 1$. 

In addition to the effect discussed above,
a finite current $J_x$ may heat the electrons
and can also affect the longitudinal resistivity through the electron-temperature 
change in the $\Pi_2$ function. Giving the drift velocity $v$ 
or the current $J_x=N_{\rm s}e v$, 
the electron temperature $T_{\rm e}$ is easily determined  
by the energy-balance equation (\ref{eqenergy}), $w(v)+vf(v)=0$,
and the longitudinal resistivity is then obtained directly from Eq.\,(\ref{eqrxx}).

For GaAs-based high-mobility 2DESs at low temperatures, 
the dominant direct contribution
to the resistivity comes from impurity scatterings and $f_{\rm p}$ is negligible.  
To obtain the electron energy dissipation rate $w$ 
needed for determining the electron heating, we consider 
scatterings from bulk longitudinal and transverse acoustic  
phonons, as well as from polar optical  
phonons with coupling parameters taken 
as typical values of $n$-type GaAs,\cite{Lei851} having an
electron effective mass $m=0.067\,m_{\rm e}$ ($m_{\rm e}$ is the
free electron mass).

\begin{figure}
\includegraphics [width=0.4\textwidth,clip=on] {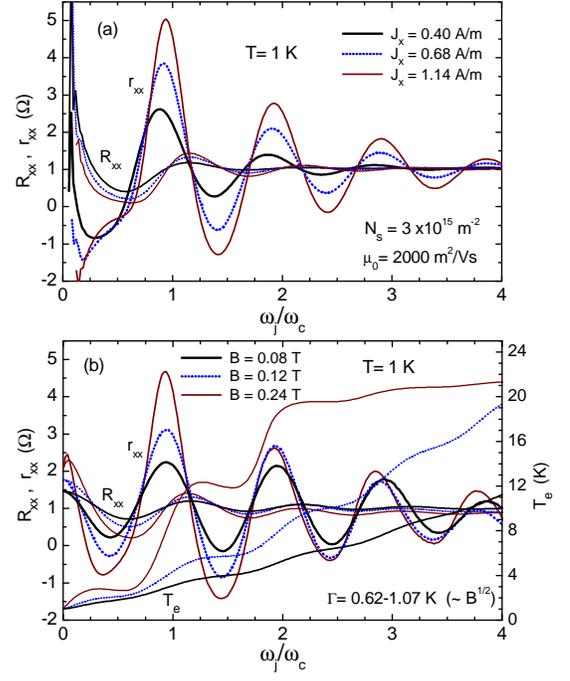}
\vspace*{-0.2cm}
\caption{(Color online) (a) Resistivity $R_{xx}$ and differential resistivity $r_{xx}$ 
 versus $\omega_j/\omega_c$ at fixed DC current density 
 $J_x=0.40, 0.68$ or 1.14\,A/m, and (b) $R_{xx}$, $r_{xx}$ 
and electron temperature $T_{\rm e}$ versus $\omega_j/\omega_c$ 
at fixed magnetic field $B=0.08,0.12$ or 0.24\,T, for
a GaAs-based 2D system having electron density
$N_{\rm s}=3.0\times 10^{15}$\,m$^{-2}$, linear mobility $\mu_0=2000$\,m$^2$/Vs,
$\alpha=10$ at $T=1$\,K, assuming SR scattering.
}
\label{fig1}
\end{figure}
\begin{figure}
\includegraphics [width=0.4\textwidth,clip=on] {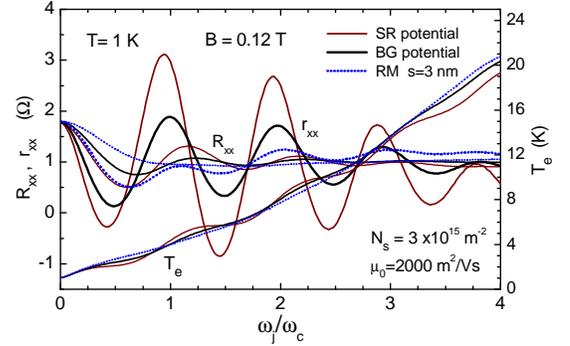}
\vspace*{-0.2cm}
\caption{(Color online) $R_{xx}$, $r_{xx}$ and $T_{\rm e}$ versus $\omega_j/\omega_c$ 
at fixed magnetic field $B=0.12$\,T for the same system as described 
in Fig.\,1 but subject to SR, BG or RM 
scattering potentials.}
\label{fig2}
\end{figure}

Figure 1(a) presents the calculated resistivity $R_{xx}$ and differential resistivity 
$r_{xx}$ versus the inverse magnetic field $1/B$ in terms of $\omega_j/\omega_c$ 
at lattice temperature  $T=1$\,K, 
for a 2D system with electron density $N_{\rm s}=3.0\times 10^{15}$\,m$^{-2}$
and linear mobility $\mu_0=2000$\,m$^2$/Vs 
subject to three different bias dc current densities 
$J_x=0.40, 0.68$ and 1.14\,A/m, which correspond to $\omega_j/2\pi=32.3, 62.2$ 
and 103.6\,GHz respectively.
The elastic scattering is assumed to be short ranged, 
and the broadening parameter $\alpha=10$, 
i.e. $B^{1/2}$-dependent $\Gamma$, $\simeq 0.62$\,K at $B=0.08$\,T. 
Oscillations in resistivity $R_{xx}$, especially in differential resistivity $r_{xx}$ 
show up remarkably, having an approximate period $\Delta(\omega_j/\omega_c)\sim 1$. 
The oscillation amplitude decays with increasing $\omega_j/\omega_c$ 
(reducing $B$ field, due to increasing overlap of LLs) at fixed bias current 
but increases with increasing bias current density within the $J_x$ range shown. 
The maxima (minima) of the differential resistivity $r_{xx}$ locate quite close 
to (but somewhat lower than) the integers (half integers) of $\omega_j/\omega_c$,
while the maxima (minima) of the total resistivity $R_{xx}$ are shifted around
a quarter period higher. These features are in good agreement with recent
experimental findings.\cite{WZ0608727}
Note that the electron temperature $T_{\rm e}$ (not shown) 
exhibits only a weak variation with changing $B$ field 
at each fixed dc.

In Fig.\,1(b) we plot $R_{xx}$, $r_{xx}$ and 
electron temperature $T_{\rm e}$ versus the dc density 
in terms of $\omega_j/\omega_c$ at fixed magnetic fields 
$B=0.08,0.12$ and 0.24\,T for the same system. 
Remarkable $R_{xx}$ and $r_{xx}$ oscillations with approximate period
$\Delta(\omega_j/\omega_c) \sim 1$ and maxima (minima) positions similar to those in 
Fig.\,1(a) can be seen in this current-sweeping figure but here
the oscillation decay with increasing $\omega_j/\omega_c$ is 
due to increase in the electron temperature.
At lower $\omega_j/\omega_c$ range, the oscillation amplitude of high 
$B$-field case is apparently larger than low $B$-field case when the electron temperature
$T_{\rm e}$ is still in the range less than or around $10$\,K. 
However, in the case of $B=0.24$\,T the oscillation amplitude decays rapidly 
with increasing $\omega_j/\omega_c$ due to the rapid increase of
electron temperature, which rises up to $20$\,K range around 
$\omega_j/\omega_c=2$. In comparison, the amplitude decay  
 is much slower in the case of $B=0.08$\,T 
because of the slow $T_{\rm e}$ rise. Furthermore, the periods 
of $R_{xx}$ and $r_{xx}$ oscillations are also somewhat shrunk by the rise 
of $T_{\rm e}$, as can be seen at higher orders in the $B=0.24$\,T case. 

\begin{figure}
\includegraphics [width=0.45\textwidth,clip=on] {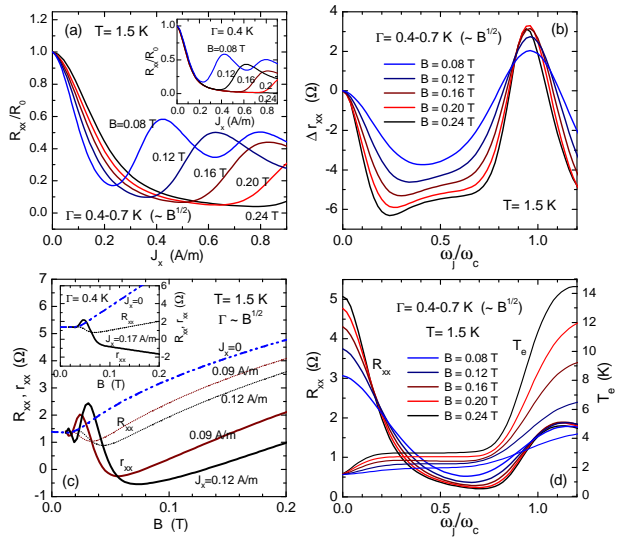}
\vspace*{-0.2cm}
\caption{(Color online) Calculated 
$R_{xx}$, $r_{xx}$, $R_{xx}/R_{0}$, $\Delta r_{xx}=r_{xx}-R_{0}$ and $T_{\rm e}$ 
for a GaAs-based 2D system with $N_{\rm s}=3.0\times 10^{15}$\,m$^{-2}$ 
at lattice temperature $T=1.5$\,K subject to a mixed SR and BG elastic scattering. 
The LL width is taken to be
$B^{1/2}$ dependent with $\Gamma=0.4$\,K at $B=0.08$\,T.
The insects illustrate the case of fixed LL width $\Gamma=0.4$\,K.}  
\label{fig3}
\end{figure} 

Note that though the periods of these resistance oscillations 
are roughly the same in terms of $\omega_j/\omega_c$, 
their amplitude and the detailed behavior depend strongly 
on the form of the scattering  
potential $U({\bf q}_{\|})$ in Eq.\,(\ref{forcei}).
To have an idea of this scattering potential effect we plot, in Fig.\,2,   
$R_{xx}$, $r_{xx}$, and $T_{\rm e}$ as functions of $\omega_j/\omega_c$
at a fixed magnetic field $B=0.12$\,T for the same 2D system
but with the dominant elastic scatterings, 
respectively, due to short-range (SR) disorder, 
charged impurities in the background (BG), 
or ionized impurities locating a distance $s=3$\,nm away from the 2D
sheet (RM).\cite{Lei851} $R_{xx}$ and $r_{xx}$ exhibit the strongest oscillations
in the case of SR potential, with a feature that the second minimum 
of $r_{xx}$ goes deeper into negative direction than the first minimum. 
In the case of BG scattering $R_{xx}$ and $r_{xx}$ oscillations, 
though weaker than those of SR-scattering case, still appear quite substantial
and the first minimum of $r_{xx}$ turns out to be the deepest one of all minima.
In the case of RM ($s=3$\,nm) scattering, these current-induced resistance 
oscillations, though existing, become much weaker than those of SR and BG scatterings.

Fig.\,3 presents the calculated 
$R_{xx}$, $r_{xx}$, $R_{xx}/R_{0}$, $\Delta r_{xx}=r_{xx}-R_{0}$ ($R_{0}$ 
is the resistivity at zero dc bias) and $T_{\rm e}$, 
for another GaAs-based 2D system 
with $N_{\rm s}=3.0\times 10^{15}$\,m$^{-2}$, 
$\mu_0=1500$\,m$^2$/Vs at $T=1.5$\,K,
focusing on the first oscillation period of $\omega_j/\omega_c$.  
The elastic scatterings are assumed due to a mixture of SR and BG 
impurities (with 2:1 contribution ratio to the linear mobility).\cite{Lei851} 
The broadening parameter is taken to be $\alpha=3$ 
($B^{1/2}$-dependent $\Gamma$, $\simeq 0.4$\,K at $B=0.08$\,T),
a somewhat smaller LL width to keep the system mainly in the separated LL regime.
The insets of Figs.\,3a and 3c show the results 
using a $B$-independent $\Gamma=0.4$\,K.
All the main features found in the experiment in the case of separated Landau 
levels,\cite{WZ0608727}
e.g., the dramatic initial suppression of the magnetoresistivity with increasing
dc density,
the widening and deepening of the first minimum range with increasing $B$ field,
a few resolvable oscillations of $r_{xx}$ at lower $B$, 
dramatic reduction of it at higher magnetic field, etc.,
are well reproduced. Note that the width of the half zero-bias peak in terms of 
$\omega_j$ accurately reflects the width of LLs, as can be seen clearly   
in Fig.\,3(a) and its inset. This suggests an ideal way to determine the width
of LLs.

This work was supported by the projects of National Science Foundation of China
and Shanghai Municipal Commission of Science and Technology.

\end{document}